\documentclass[letterpaper,11pt]{article} 

\usepackage{amsmath, amssymb, amsthm, braket, enumitem, dsfont, relsize}
\usepackage{opticameet3} 
\usepackage{setspace}
\usepackage[utf8]{inputenc}
\usepackage[T1]{fontenc}
\usepackage{lineno}
\usepackage[backend=biber, style=nature, articletitle=true, date=year]{biblatex}
\newcommand\authormark[1]{\textsuperscript{#1}}

\usepackage{amsmath,amssymb}
\usepackage{caption}
\usepackage{svg}
\usepackage[colorlinks=true,bookmarks=false,citecolor=blue,urlcolor=blue]{hyperref} 

\begin{document}


\DeclareFieldFormat[inproceedings,proceedings]{title}{#1}

\title{Link-Free Multi-Node Timing Synchronization for Scalable Quantum Networking}


\author{
Jacob E. Humberd,\authormark{1}
Mohmad Junaid Ul Haq,\authormark{2,3 *}
Angel Fraire Estrada,\authormark{2,3}
\newline
Ike Deitch,\authormark{2,3}
Tian Li\authormark{1 †}
}


\address{
\authormark{1}Department of Electrical \& Computer Engineering, Florida State University, Tallahassee, FL 32306, USA\\
\authormark{2}Department of Physics \& Astronomy, University of Tennessee, Chattanooga, TN 37403, USA\\
\authormark{3}UTC Quantum Center, University of Tennessee, Chattanooga, TN 37403, USA\\
}

\email{\authormark{*}mohmad-junaidulhaq@utc.edu; \authormark{†}tli6@fsu.edu}

\begin{abstract*}
Precise timing synchronization is a fundamental requirement for distributed quantum networking, enabling the identification of nonlocal photon correlations underlying entanglement distribution, quantum teleportation, and entanglement swapping. Existing synchronization architectures rely on dedicated timing-distribution infrastructure, most notably White Rabbit networks, which impose constraints on topology, scalability, and deployment in free-space and satellite environments. Here we demonstrate link-free synchronization of quantum network nodes using independently operating miniature rubidium atomic clocks combined with computational post-processing. A two-stage coincidence-peak search algorithm recovers photon correlations despite non-deterministic acquisition offsets exceeding one second, while a drift-compensation procedure corrects the accumulated phase evolution between independent clocks over multi-hour operation. We validate the approach on a deployed metropolitan-scale telecom fiber network spanning 3 geographically separated nodes. Following drift correction, our atomic-clock-based synchronization achieves timing performance approaching that of a White Rabbit benchmark and remains stable over continuous 8-hour measurement intervals. As a stringent test of quantum networking functionality, we observe Hong–Ou–Mandel interference across the spatially separated nodes with visibility exceeding 70\%, statistically equivalent to that obtained using dedicated White Rabbit timing-distribution links. To the best of our knowledge, this constitutes the first observation of quantum interference across a deployed metropolitan-scale telecom fiber network synchronized entirely without dedicated timing-transfer infrastructure. These results establish our atomic-clock-based synchronization as a scalable, topology-independent alternative to conventional timing-distribution architectures and provide a practical pathway toward terrestrial, airborne, and space-based quantum networks where dedicated timing links are unavailable.
\\
\end{abstract*}

\section{Introduction}
Deployed quantum networks require geographically separated nodes to share a common time reference with sub-nanosecond accuracy and pico-second precision. This capability is fundamental to a wide range of quantum networking protocols, including entanglement distribution \supercite{1,2,3}, entanglement swapping \supercite{4,5,6}, measurement-device-independent quantum key distribution (QKD) \supercite{7,8,9,10}, and distributed quantum metrology such as networks of entangled clocks \supercite{11}. In each of these applications, the ability to assign accurate and mutually consistent timestamps to photon detection events across remote locations is essential for identifying genuine quantum correlations and distinguishing them from the background of accidental coincidences. Without precise synchronization, the coincidence window must be broadened, leading to increased noise background, reduced coincidence-to-accidental ratio (CAR), and degraded network performance \supercite{12,13}.

The establishment of a shared time reference across remote nodes requires addressing two distinct problems. The first is \textit{synchronization}, defined as the alignment of the absolute time origins of clocks that may have been initialized independently. The second is \textit{syntonization}, defined as the matching of oscillator frequencies to ensure a common rate of time progression. While synchronization establishes a common temporal reference, syntonization is required to maintain it. 
Two clocks aligned to a common time origin but whose oscillators differ in rate by even a part in 10$^{11}$ will steadily accumulate a relative timing error, growing in proportion to the elapsed time.
For the independent-running rubidium frequency standards used in this work, the 10$^{-11}$ fractional frequency error can generate timing errors on the order of $20~\mathrm{ps/s}$ in the holdover mode, which accumulate into nanosecond-scale timing offsets over the course of a typical multi-minute acquisition.
\color{black}
For coincidence measurements spanning to hours, both the initial synchronization offset and the accumulated timing drift arising from residual frequency differences must be accurately estimated and compensated to maintain stable, chromatic dispersion-limited photon correlation profiles. 

\color{black}
The current widely adopted solution is the White Rabbit Precision Timing Protocol (WRPTP) \supercite{16,17,18}, which distributes a common clock reference over fiber-optic links and actively disciplines the local oscillators of connected devices through closed-loop feedback control. By simultaneously providing synchronization and syntonization, WRPTP achieves sub-nanosecond timing accuracy and continuously compensates for oscillator drift, maintaining tens of picoseconds scale timing precision in deployed telecom fiber networks \supercite{13,19,20}.
Despite its high timing accuracy, WRPTP is subject to a fundamental architectural constraint: every synchronized node must maintain a dedicated fiber connection to a central White Rabbit switch (WRS), effectively enforcing a star-topology network architecture \supercite{21,22}. This requirement limits scalability by increasing per-node fiber overhead and restricts deployment to environments with fixed fiber infrastructure. Consequently, WRPTP is not readily applicable to free-space quantum links, airborne platforms, satellite-based quantum networks where dedicated timing links are unavailable \supercite{3,23}. It is also worth noting that a complementary class of approaches eliminates the need for dedicated timing hardware by syntonizing remote rubidium clocks using the temporal correlations of photon pairs distributed over the link \supercite{pelet}. Unlike the approach presented here, these methods still rely on a continuously active timing
transfer channel to maintain synchronization.


Miniature atomic clocks (MACs) provide a connectivity-independent synchronization approach for distributed quantum networks. Rubidium-based atomic frequency standards exhibit sufficient stability to maintain a shared time reference over multi-hour operational periods. By synchronizing two clocks at a common location prior to deployment, timing synchronization can be transferred to geographically separated nodes without the need for dedicated timing links \supercite{12,24,25}. While residual frequency offsets lead to gradual divergence between the clocks, the resulting timing drift is slow and deterministic, and can therefore be readily characterized, enabling accurate compensation through post-processing calibration.

From a resource perspective, MAC-based synchronization eliminates the dedicated timing-fiber infrastructure altogether \supercite{21,22}. While a locally deployed MAC provides the timing reference, each node requires only its quantum signal channels, thereby reducing deployment complexity and improving network scalability. This architecture significantly reduces per-node fiber overhead relative to WRPTP and removes the centralized-star topology requirement, enabling synchronization across distributed nodes without direct connections to a central WRS. 
The resulting architecture is therefore topology-agnostic, supporting arbitrary network configurations without modification \supercite{3,22}, and readily extends to heterogeneous communication links, including free-space and satellite channels, where dedicated timing fibers cannot be deployed \supercite{3,23}.

\color{black}

A further challenge unique to multi-node quantum measurements is the unknown and run-to-run varying timing offset introduced by the software initialization procedures of commercial time-tagging hardware.
When independent time-tagging modules are initialized at geographically separated locations, the proprietary acquisition interface provides no deterministic control over the precise onset of data acquisition. Consequently, an unknown inter-file timing offset, ranging from hundreds of milliseconds to more than one second, is introduced between the recorded datasets. Because this offset is several orders of magnitude larger than the physical propagation delay between the quantum channel signals, direct cross-correlation of the raw timestamp files becomes computationally prohibitive.

\color{black}
We report connection-free synchronization of quantum network nodes using independent-running miniature rubidium atomic clocks (MACs) combined with computational post-processing. Our approach employs a two-stage coincidence peak search algorithm to efficiently recover photon correlations despite large unknown timing offsets, and a piecewise linear drift compensation method to correct the accumulated divergence arising from residual frequency differences between independent oscillators. We experimentally validate the technique across geographically distributed nodes on a deployed metropolitan-scale fiber quantum network in downtown Chattanooga, Tennessee. Performance is benchmarked against the WRS timing reference over continuous measurement intervals exceeding 8 hours, demonstrating the viability of connectivity-independent timing synchronization for scalable quantum networks.

\color{black}
We further validate the quantum utility of the our approach by observing Hong–Ou–Mandel (HOM) two-photon interference among 3 geographically separated network nodes synchronized solely by MACs. The observed visibility exceeds 70~\%, demonstrating our connection-free synchronization is viable to  support high-visibility quantum interference across deployed fiber infrastructure. Together, these results establish a practical, scalable, and connectivity-independent synchronization framework for future quantum networks, including free-space, airborne, satellite-based systems where dedicated timing links cannot be readily established.

\begin{figure}[t]
  \centering
  \includegraphics[width=1\linewidth]{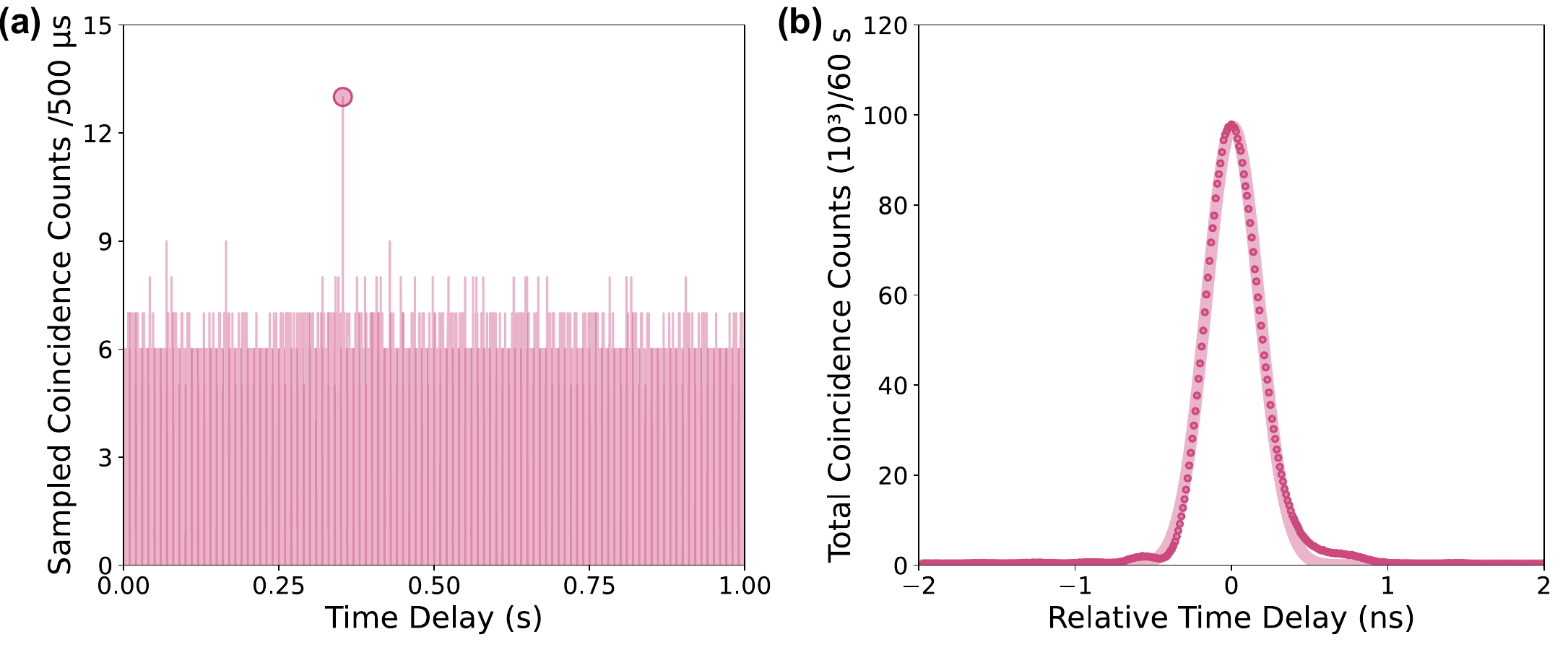}
  \caption{Experimental results of the coarse- and fine-stage coincidence peak identification algorithm. (a) A coarse search using temporally subsampled timestamp streams and a 10~ns time-bin width localizes the approximate position of the coincidence peak. (b) The full timestamp records are then analyzed in the vicinity of the identified peak using a 10~ps time-bin width to reconstruct the high-resolution distribution of the second-order cross-correlation function $G^{(2)}(\tau)$. The solid pink line denotes a Gaussian fit to the measured correlation profile.
  }
\label{fig:scan}
\end{figure}

\section{Results}

\subsection{Coincidence peak identification from distributed timestamp streams}

A key challenge in multi-node quantum coincidence measurements arises from the fact that independently operated time-tagging modules, each controlled by a separate computer at a remote location, generate timestamp files whose relative temporal origins are inherently unknown. Variability in the software initialization process of the proprietary acquisition interface also introduces a non-deterministic offset between datasets. This timing offset can exceed the physical photon propagation delay between quantum channels by 3 to 6 orders of magnitude, dramatically expanding the coincidence search space and rendering brute-force evaluation of the second-order cross-correlation function, $G^{(2)}(\tau)$, computationally prohibitive.

Throughout this work, photon coincidences were generated using a spatially degenerate continuous-wave (CW) bi-photon source (BPS) based on type-II spontaneous parametric down-conversion (SPDC) in a periodically poled KTP (PPKTP) crystal. The crystal temperature was adjusted to generate photon pairs centered near 1570~nm with an approximate bandwidth of 3~nm. The orthogonally polarized photons in each pair were separated by a polarizing beam splitter (PBS). To maximize detection efficiency, the polarization-maintaining (PM) fiber connected to the reflected PBS output was rotated so that both photons were coupled into the same horizontal polarization state before detection by superconducting nanowire single-photon detectors (SNSPDs).

\begin{figure}[t]
  \centering
  \includegraphics[width=1\linewidth]{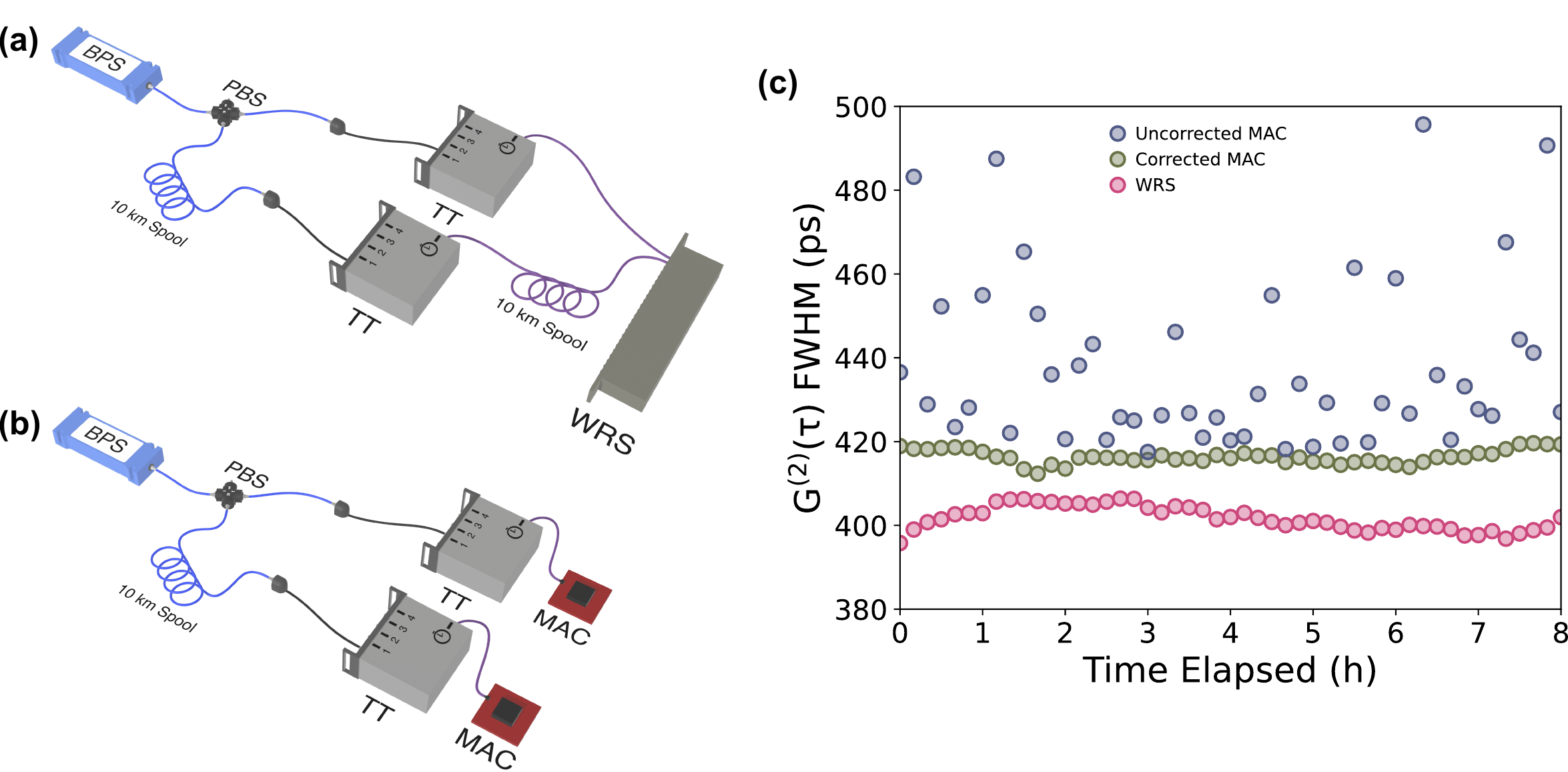}
  \caption{Local experimental configurations at UTC employing 10~km fiber spools for comparison of the synchronization performance using (a) WRS-based and (b) MAC-based method. (c) Measured coincidence peak FWHM as a function of acquisition time over an 8-hour measurement session using WRS (pink), uncorrected MACs (blue), and drift-corrected MACs (green). The coincidences were sampled at 10-minute intervals, with each data point corresponding to a 60-second integration period.}
\label{fig:localcorrelation}
\end{figure}

We addressed the coincidence peak identification challenge using a two-stage coincidence peak search algorithm that localizes the correlation peak without evaluating the full correlation function across every candidate offset. As shown in \textbf{Fig.~\ref{fig:scan}(a)}, the coarse stage employs a deliberately large time-bin width of 10~ns to obtain a rough estimate of where the correlation peak lies within the timestamp stream. Because only selected portions of each file are searched at this stage rather than the complete dataset, the sampled coincidence counts are correspondingly low, yet sufficient to identify the approximate peak location even for acquisition start time offsets exceeding 1s. The fine stage then rebuilds the correlation around this identified location using the full timestamp records, reconstructing the $G^{(2)}(\tau)$ distribution at high resolution as shown in \textbf{Fig.~\ref{fig:scan}(b)}, from which the peak center and full width at half maximum (FWHM) are extracted via Gaussian fitting. The algorithm successfully recovered the correlation peak in all experimental configurations, including cases where the acquisition start time offset exceeded 1s.

\subsection{Synchronization of distributed time-tagging modules: local configuration with fiber spools}

Leveraging the coincidence peak identification algorithm described above, we first characterize the synchronization performance of distributed time-tagging modules under controlled laboratory conditions. In this configuration, both time-tagging modules were co-located in the UTC Quantum Node Lab, while 10~km fiber delay spools were incorporated to emulate the path lengths encountered in deployed quantum networks. Three synchronization scenarios were evaluated: WRS synchronization using the experimental setup shown in \textbf{Fig.~\ref{fig:localcorrelation}(a)}, and both independent-running and drift-corrected MAC synchronization using the  experimental setup shown in \textbf{Fig.~\ref{fig:localcorrelation}(b)}. The FWHM of the coincidence peak was tracked as a function of elapsed time over an 8-hour measurement session, serving as the primary figure of merit for synchronization performance. 

\begin{figure}[t]
  \centering
  \includegraphics[width=1\linewidth]{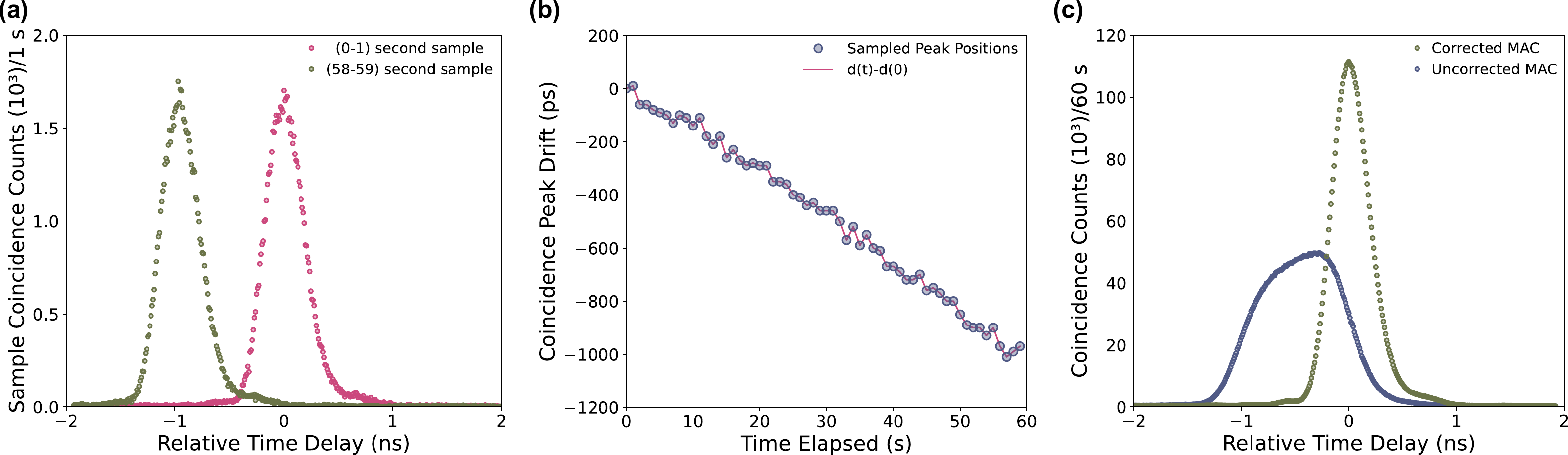}
  \caption{(a) Comparison of the first (pink) and last (green) one-second intervals of a 1-minute acquisition using independent MACs, illustrating the displacement of the coincidence peak caused by accumulated timing drift between the MACs. (b) Coincidence peak position as a function of acquisition time over the same 1-minute interval. Each data point corresponds to the peak location extracted from a one-second segment of the timestamp stream. The solid pink line represents the time-dependent drift model constructed using piecewise linear fit. (c) Comparison of the reconstructed $G^{(2)}(\tau)$ distributions obtained from the uncorrected MAC data (blue) and the same dataset after application of the drift-correction model (green).
  }
\label{fig:driftcorrection}
\end{figure}

Under the WRS synchronization, the FWHM remained essentially constant over the full 8-hour measurement session, averaging $401.7 \pm 2.8~\mathrm{ps}$, as shown by the pink dots in \textbf{Fig.~\ref{fig:localcorrelation}(c)}. The observed stability results from the closed-loop phase-locking mechanism implemented by the WRS. By referencing the local time-base of each time tagger to a common master 10~MHz distributed by the WRS, the closed phase-locking loop  continuously suppresses the relative phase evolution between the two clocks, which is sufficient to maintain a stable coincidence FWHM. Whereas, under the independent-running atomic clock synchronization, the FWHM increased progressively over time, reflecting the cumulative relative phase drift between the two independent 10 MHz outputs as shown by the blue dots in \textbf{Fig.~\ref{fig:localcorrelation}(c)}. 
In each MAC, a 10~MHz oven-controlled crystal oscillator (OCXO) is frequency-disciplined to the Rubidium-87 hyperfine transition through a feedback voltage control loop. The measured drift therefore originates from the residual frequency mismatch between the two independently disciplined quartz oscillators, which accumulates as a relative phase offset over time, rather than from any appreciable variation of the atomic reference itself. The accumulated timing drift is directly visualized in \textbf{Fig.~\ref{fig:driftcorrection}(a)}, which compares the coincidence peaks measured during the first and last one-second intervals of 1-minute acquisition period. The pronounced displacement of the correlation peak reflects the continuous phase accumulation between the two clocks.

To quantify the temporal evolution of the phase drift, each timestamp file was segmented into one second intervals, and the peak position of the $G^{(2)}(\tau)$ distribution profile was extracted for each interval. The resulting time series provides a direct measure of the accumulated phase drift over the duration of the measurement as shown in \textbf{Fig.~\ref{fig:driftcorrection}(b)}. A piecewise linear drift model was constructed from the sequence of per-second distribution profile peak positions, and linear interpolation was used to estimate the drift at arbitrary timestamps, thereby generating a continuous correction function $d(t)$. After applying the time-dependent correction
$t' = t - \bigl(d(t)-d(0)\bigr)$, 
to one of the timestamp streams, the reconstructed FWHM of the $G^{(2)}(\tau)$ distribution was reduced to $416.4 \pm 1.9~\mathrm{ps}$, approaching the WRS benchmark of $401.7 \pm 2.8~\mathrm{ps}$. A representative comparison between the corrected and uncorrected MAC data is shown in \textbf{Fig.~\ref{fig:driftcorrection}(c)}. The corrected performance remained stable throughout the full 8-hour acquisition period, as indicated by the green dots in \textbf{Fig.~\ref{fig:localcorrelation}(c)}. These results demonstrate that the dominant timing error arises from the accumulated relative phase drift between the independent-running rubidium frequency standards and can be effectively removed through our post-processing algorithm.





\subsection{Synchronization of distributed time-tagging modules: non-local configuration with deployed fibers}

\begin{figure}[t]
  \centering
  \includegraphics[width=1\linewidth]{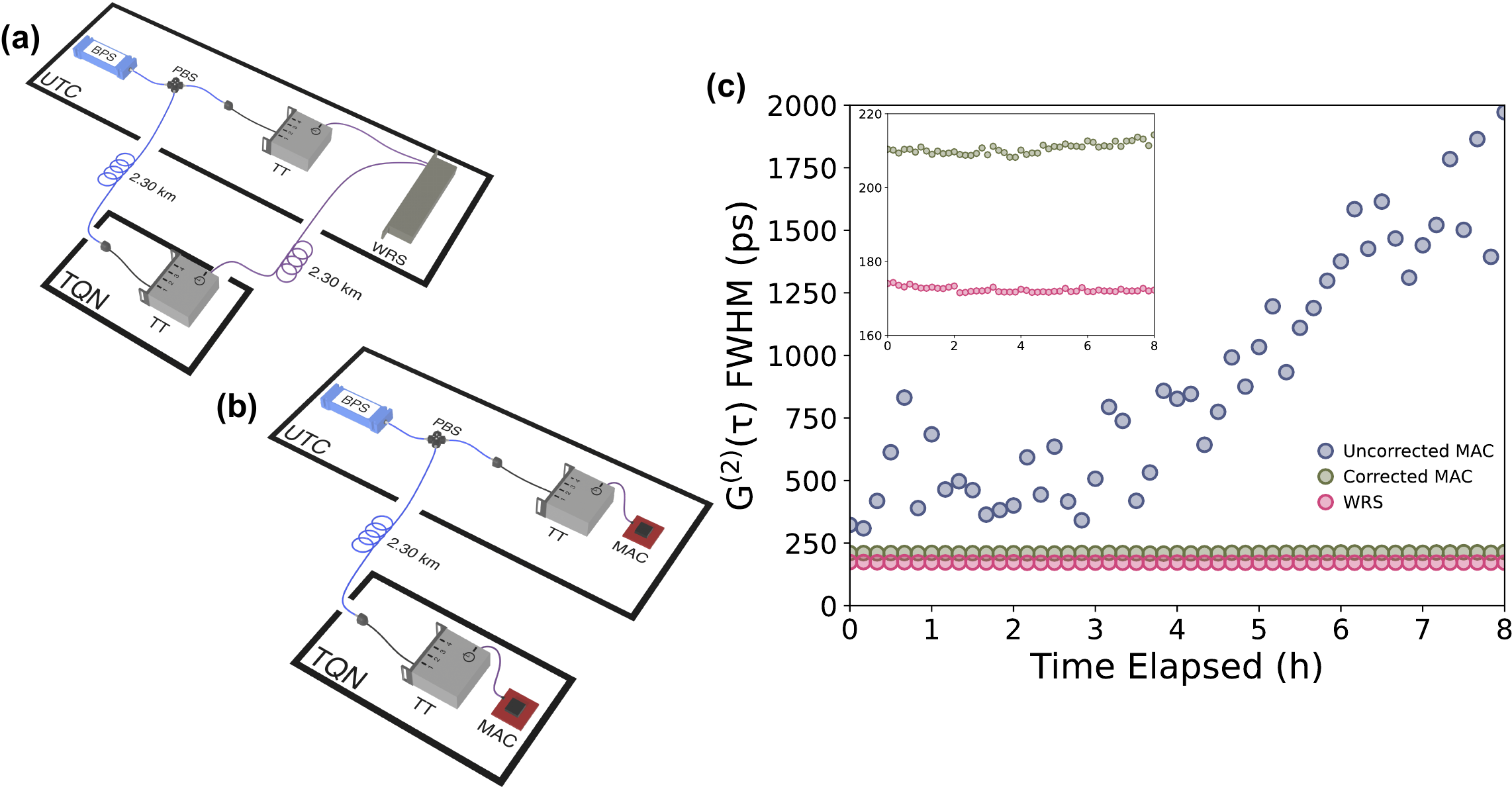}
  \caption{Deployed network experimental configurations connecting the UTC node and the Tenth Street Quantum Node (TQN) for comparison of the synchronization performance using (a) WRS-based and (b) MAC-based method. (c) Measured coincidence peak FWHM as a function of acquisition time over an 8-hour measurement session using WRS (pink), uncorrected MACs (blue), and drift-corrected MACs (green). The coincidences were sampled at 10-minute intervals, with each data point corresponding to a 60-second integration period. The deployed telecom fiber link between UTC and TQN is 2.30~km in length. The inset in (c) presents the same data on an expanded scale for clarity.
  }
\label{fig:networkcorrelation}
\end{figure}

Having established synchronization performance under controlled laboratory conditions, we next evaluate our scheme on deployed fiber infrastructure. One time-tagger was deployed at the UTC Quantum Node Lab and the other at the Tenth Street Quantum Node (TQN), with the two sites connected via dedicated 2.3~km telecommunications fiber links. The corresponding WRS- and MAC-synchronized configurations are shown in \textbf{Fig.~\ref{fig:networkcorrelation}(a)} and \textbf{Fig.~\ref{fig:networkcorrelation}(b)}, respectively. The synchronization performance was again quantified by tracking the FWHM of the coincidence peak throughout an 8-hour acquisition, and using it as the primary figure of merit.

Under the WRS synchronization, the deployed fiber network yielded an average FWHM of $172.4 \pm 0.7~\mathrm{ps}$, as indicated by the pink dots in \textbf{Fig.~\ref{fig:networkcorrelation}(c)}. This value is substantially smaller than the $401.7 \pm 2.8~\mathrm{ps}$ observed in the local fiber-spool configuration as shown by the pink dots in \textbf{Fig.~\ref{fig:localcorrelation}(c)}. This reduction in FWHM can be largely attributed to the reduced group velocity dispersion (17~ps/nm/km at 1550~nm) in the 2.3~km deployed fiber link relative to the 10~km laboratory fiber-spool setup.

Under the independent-running MAC synchronization, the FWHM increased substantially throughout the 8-hour acquisition, reaching $1971.9~\mathrm{ps}$ by the end of the measurement period, as indicated by the blue dots in \textbf{Fig.~\ref{fig:networkcorrelation}(c)}. The observed broadening corresponds to an accumulated timing drift of approximately $2~\mathrm{ns}$ over the measurement interval. We attribute the larger drift to the delay between MAC initialization (see Methods Section 4.3) and data acquisition. After being initialized overnight in the UTC Quantum Node Lab, the MACs were deployed to the remote nodes and operated in an independent-running mode until measurements began approximately 12 hours later. Consequently, residual frequency mismatches between the two clocks accumulated as a significant relative phase offset before the acquisition commenced. This sequence reflects a realistic deployment scenario involving overnight initialization/calibration followed by daytime operation. 

By applying the same per-second drift characterization and piecewise linear correction procedure described above, the FWHM was reduced from $1971.9~\mathrm{ps}$ to $210.6 \pm 1.6~\mathrm{ps}$, approaching the WRS benchmark of $172.4 \pm 0.7~\mathrm{ps}$. The corrected coincidence peak remained stable over the full 8-hour acquisition period, as shown by the green dots in \textbf{Fig.~\ref{fig:networkcorrelation}(c)}, where a zoomed-in view is also provided in the inset. At the end of the 8-hour period, our drift correction algorithm restored the FWHM from $1971.9~\mathrm{ps}$ to $210.6~\mathrm{ps}$, yielding an approximately 9-fold improvement and recovering performance close to the WRS benchmark. These results demonstrate that our drift correction method remains effective despite large accumulated phase offsets and the environmental variability characteristic of deployed metropolitan fiber infrastructure.

\begin{figure}[t]
  \centering
  \includegraphics[width=1\linewidth]{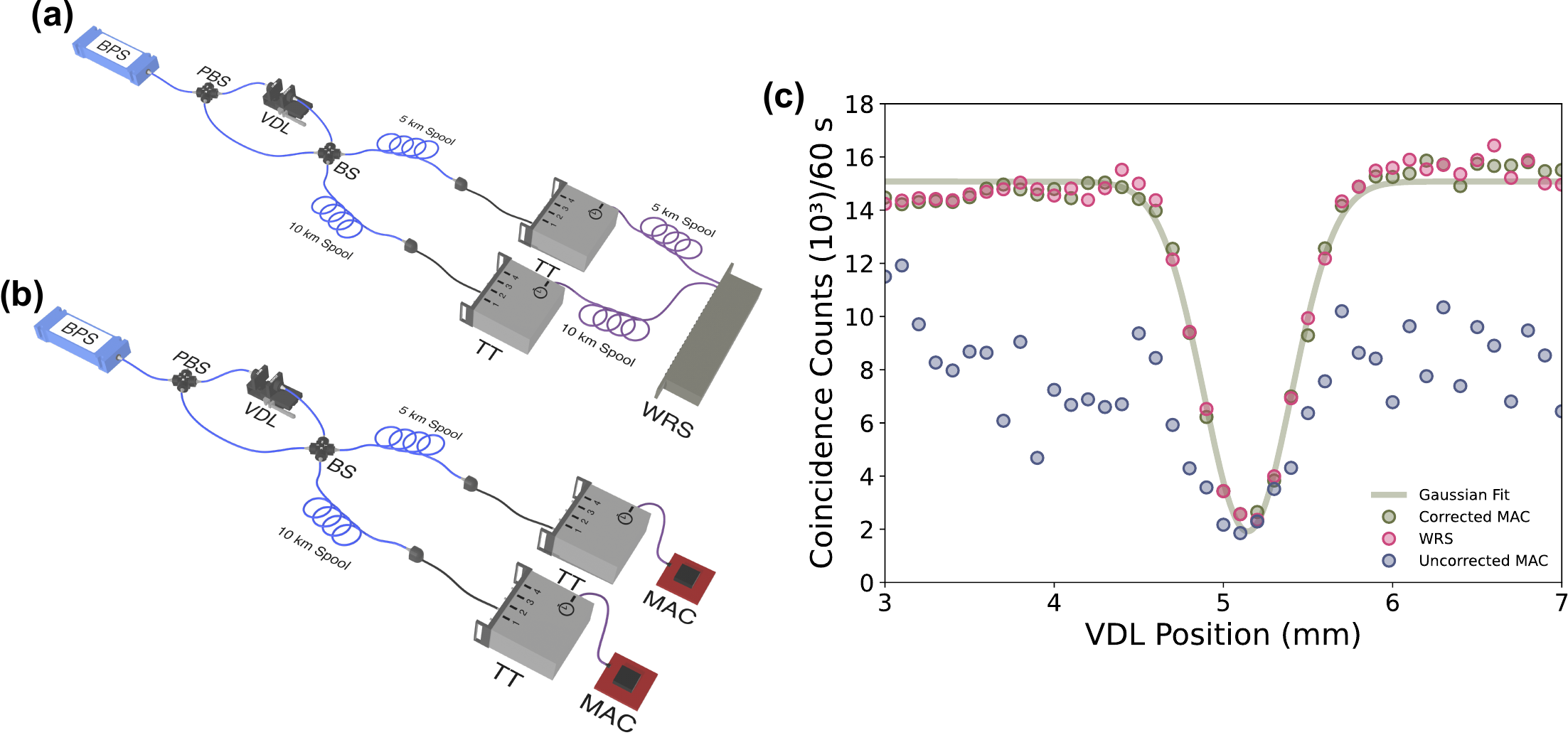}
  \caption{Local experimental configurations at UTC employing fiber spools for comparison of the HOM interference performance using (a) WRS-based and (b) MAC-based method. (c) Measured HOM dips using WRS (pink), uncorrected MACs (blue), and drift-corrected MACs (green).  The solid green line denotes a Gaussian fit to the measured HOM dip profile.
  }
\label{fig:localhom}
\end{figure}

\begin{figure}[t]
  \centering
  \includegraphics[width=1\linewidth]{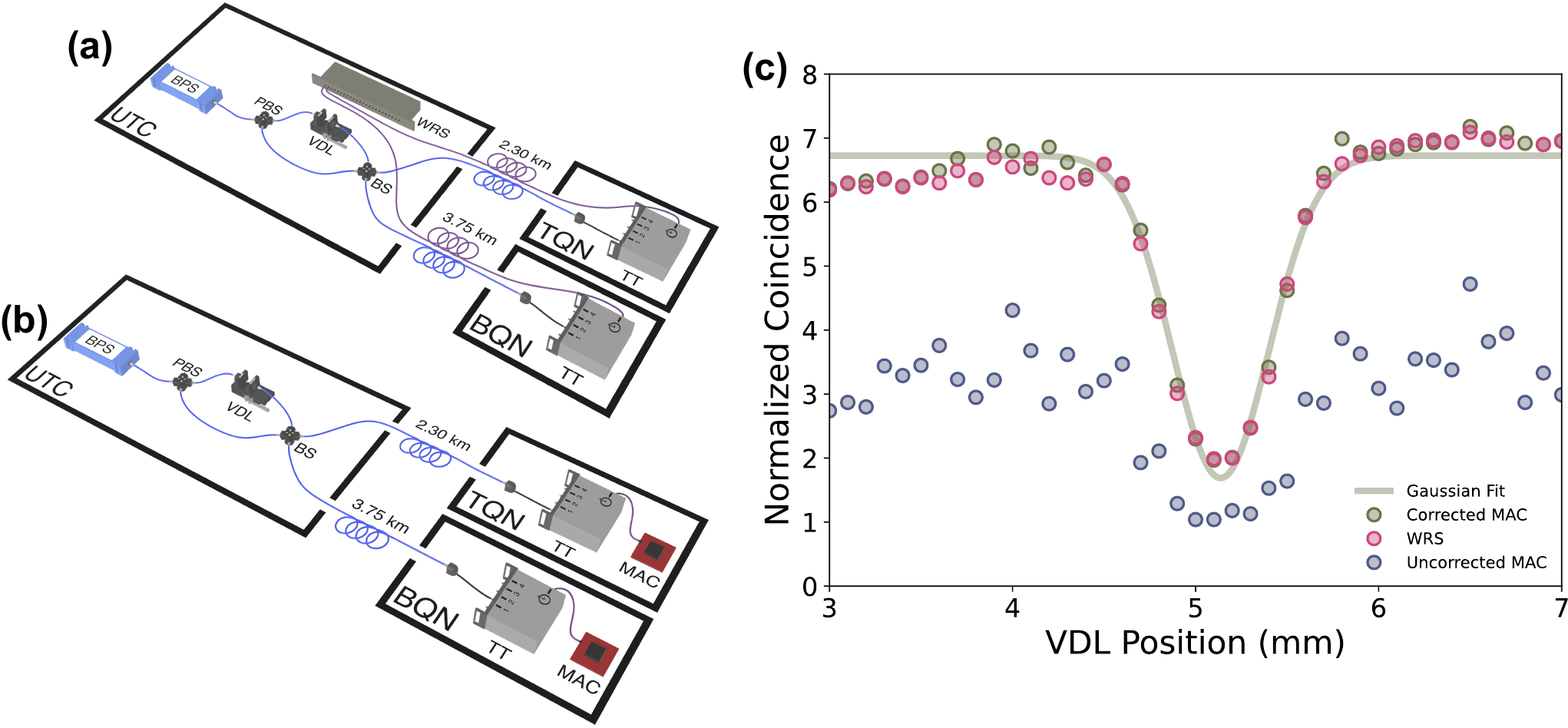}
  \caption{Deployed network experimental configurations connecting the UTC node, the Tenth Street Quantum Node (TQN), and the Broad Street Quantum Node (BQN) for comparison of the HOM interference performance using (a) WRS-based and (b) MAC-based method. The quantum interference was generated at the UTC node, detected jointly by the TQN and BQN nodes, and subsequently analyzed through post-processing at UTC. The deployed telecom fiber links between UTC and TQN and between UTC and BQN are 2.30~km and 3.75~km in length, respectively. (c) Measured HOM dips using WRS (pink), uncorrected MACs (blue), and drift-corrected MACs (green). The solid pink line denotes a Gaussian fit to the measured HOM dip profile. 
  }
\label{fig:networkhom}
\end{figure}

\subsection{Distributed quantum interference observation}

To assess the quantum networking utility of the MAC-based connection-free synchronization scheme, we conducted Hong--Ou--Mandel (HOM) interference measurements across four configurations encompassing both the WRS-based (\textbf{Figs.~\ref{fig:localhom}(a)}\&\textbf{\ref{fig:networkhom}(a)}) and the MAC-based (\textbf{Figs.~\ref{fig:localhom}(b)}\&\textbf{\ref{fig:networkhom}(b)}) synchronization methods for both laboratory (\textbf{Fig.~\ref{fig:localhom}}) and deployed (\textbf{Fig.~\ref{fig:networkhom}}) network settings. In the HOM interference measurements, the correlated photons separated by the preceding PBS were recombined at a 50:50 beam splitter (BS), with their relative arrival time controlled by a motorized variable delay line (VDL). When the photons are indistinguishable, destructive two-photon interference suppresses coincidence peaks at the BS outputs, giving rise to the characteristic HOM dip. Experimentally, for a CW source, the dip is obtained by post-processing coincidence statistics from the timestamp streams acquired at the two output ports of the BS. A visibility greater than 50\% exceeds the classical bound and serves as a clear signature of quantum interference. 
It is important to note that the coincidence peak remains the primary observable in this experiment, while the dip is obtained indirectly. Specifically, the VDL is scanned, and coincidence events are accumulated for 60 s at each VDL position. The height of the coincidence peak is then extracted and plotted as a function of the VDL position.
For both MAC-based configurations shown in \textbf{Figs.~\ref{fig:localhom}(b)\&\ref{fig:networkhom}(b)}, the drift correction was applied to the raw timestamp streams prior to the HOM analysis. By removing the accumulated relative phase drift between the two MACs, the correction preserves the HOM interference encoded in the coincidence statistics and prevents drift-induced broadening of the reconstructed HOM dip.

In the laboratory WRS configuration (\textbf{Fig.~\ref{fig:localhom}(a)}), we measured a HOM visibility of $83.6\%$ as shown by the pink dots in \textbf{Fig.~\ref{fig:localhom}(c)}, which serves as the benchmark for source performance. Following drift correction, the local MAC configuration (\textbf{Fig.~\ref{fig:localhom}(b)}) yielded a visibility of $82.6\%$ as shown by the green dots in \textbf{Fig.~\ref{fig:localhom}(c)}, statistically consistent with the WRS result. This recovery of visibility demonstrates the restoration of photon indistinguishability, in contrast to the substantially degraded visibility observed in the uncorrected case represented by the blue dots in \textbf{Fig.~\ref{fig:localhom}(c)}. The measurement was subsequently repeated on the deployed network configuration shown in \textbf{Fig.~\ref{fig:networkhom}}, where \textit{the quantum interference was generated at the UTC node, detected jointly by the TQN and BQN nodes, and subsequently analyzed through post-processing at UTC}. The deployed telecom fiber links between UTC and TQN and between UTC and BQN are 2.30~km and 3.75~km in length, respectively. Under the WRS synchronization, HOM interference was observed with a visibility of $70.0\%$, as indicated by the pink dots in \textbf{Fig.~\ref{fig:networkhom}(c)}. When synchronization was provided by the drift-corrected MACs, a clear HOM dip was recovered with a visibility of $70.2\%$  as shown by the green dots in \textbf{Fig.~\ref{fig:networkhom}(c)}, statistically equivalent to the WRS result. In contrast, the independent-running MAC-based configuration exhibited a substantially reduced visibility as indicated by the blue dots in \textbf{Fig.~\ref{fig:networkhom}(c)}, consistent with the behavior observed in the local fiber-spool configuration. Note that the coincidence counts in \textbf{Fig.~\ref{fig:networkhom}(c)} were normalized by dividing each data point by the product of the corresponding single counts in the two timestamp streams. This normalization was introduced to compensate for fluctuations in the individual timestamp streams caused by counts imbalance and environmental variations encountered during the field measurement, thereby preventing these effects from artificially influencing the observed HOM visibility.

To the best of our knowledge, this is the first demonstration of quantum two-photon interference across a deployed metropolitan-scale quantum network comprising 3 spatially separated nodes synchronized entirely without dedicated timing-distribution links. These results establish that high-visibility quantum interference can be preserved using connection-free synchronization, removing a major infrastructure constraint for future distributed quantum information systems.



\section{Discussion and Conclusion}

We have demonstrated connection-free, multi-node timing synchronization for a deployed metropolitan-scale quantum network using MACs in combination with computational post-processing. A two-stage coincidence peak identification algorithm successfully recovered photon correlations despite non-deterministic acquisition start-time offsets exceeding one second, while a piecewise linear drift correction procedure compensated the relative phase accumulation between independently operating MACs.

Under controlled laboratory conditions with 10~km fiber spools, the drift correction restored the coincidence FWHM to $416.4 \pm 1.9~\mathrm{ps}$, approaching the WR benchmark of $401.7 \pm 2.8~\mathrm{ps}$. On the deployed UTC--TQN metro-scale fiber link, a 12-hour interval between MAC initialization and data acquisition resulted in substantial drift of approximately 2 ns prior to the start of measurement, broadening the coincidence FWHM to $1971.9~\mathrm{ps}$. Nevertheless, the same correction procedure reduced the FWHM to $210.6 \pm 1.6~\mathrm{ps}$, representing an approximately 9-fold improvement and again approaching the WR benchmark of $172.4 \pm 0.7~\mathrm{ps}$. In both laboratory and deployed-network configurations, the corrected timing performance remained stable throughout continuous 8-hour acquisition periods. 


The quantum networking utility of our synchronization scheme was validated through the recovery of HOM interference between spatially separated nodes on the deployed metro-scale fiber network. Within each network configuration, the visibilities obtained under WR-based synchronization and drift-corrected MAC-based synchronization were nearly identical, yielding $83.6\%$ and $82.6\%$ respectively in the laboratory configuration, and $70.0\%$ and $70.2\%$ respectively on the deployed network configuration. These results demonstrate that the drift-correction procedure introduces no measurable degradation in photon indistinguishability. The reduction in visibility from the laboratory ($\sim83\%$) to the deployed ($\sim70\%$) configuration is attributed primarily to the environmental constraints experienced by the deployed quantum channels rather than the timing reference. In particular, uncontrolled polarization evolution along the deployed fiber links can scramble the photon polarization state prior to detection, leading to polarization-dependent variations in SNSPD detection efficiency and consequently reducing the observed HOM interference contrast. The unequal link lengths and channel losses between the TQN and BQN paths introduce count rate imbalances, which further reduce the interference contrast. Despite these network-induced impairments, the persistence of HOM visibility above the $50\%$ classical limit nonetheless confirms the preservation of nonclassical two-photon interference on the deployed fiber infrastructure. 

These results constitute the first demonstration of quantum interference across 3 distant nodes on a deployed metro-scale fiber infrastructure synchronized entirely without dedicated timing transfer links. The demonstrated compatibility of connection-free MAC-based synchronization with quantum networking protocols such as distributed entanglement distribution, quantum teleportation, and entanglement swapping establishes a viable alternative to conventional timing-transfer architectures. By eliminating the need for dedicated synchronization infrastructure while preserving the high-visibility quantum interference required for network operations, this approach provides a scalable and infrastructure-efficient pathway toward future large-scale quantum networks.
Perhaps the most compelling future direction is the extension of connection-free synchronization to free-space and satellite quantum links. Unlike terrestrial fiber networks, these architectures cannot rely on dedicated timing-distribution channels, rendering atomic clocks the only practical synchronization solution. A successful demonstration in this regime would validate the approach under the most challenging operational conditions and establish a critical enabling capability for space-to-ground quantum networking and future global-scale quantum information infrastructures.

\section{Methods}

\subsection{Deployed metro-scale fiber infrastructure}

\begin{figure}[t]
  \centering
  \includegraphics[width=1\linewidth]{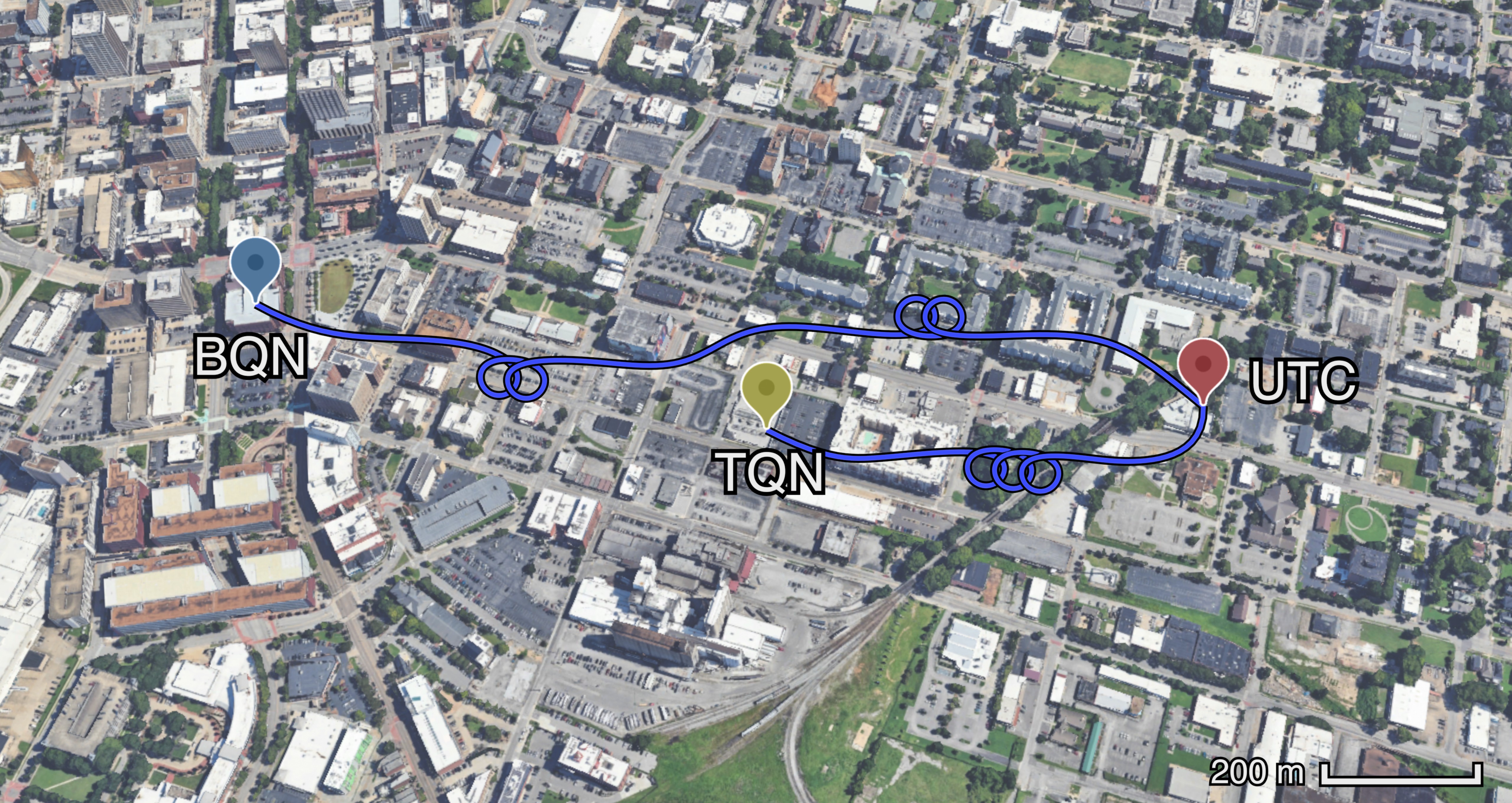}
  \caption{Bird's-eye view of the deployed telecomm fiber network in downtown Chattanooga, Tennessee, illustrating the connectivity among the UTC Quantum Node, the Tenth Street Quantum Node (TQN), and the Broad Street Quantum Node (BQN). The deployed telecom fiber links between UTC and TQN and between UTC and BQN are 2.30~km and 3.75~km in length, respectively. 
  \color{black}}
\label{fig:networkmap}
\end{figure}

All experiments were performed on the EPB fiber network in Chattanooga, Tennessee, a deployed metropolitan-scale telecommunications infrastructure~\supercite{earl2022architecture}. The network consisted of 3 nodes: the UTC Quantum Node, the Tenth Street Quantum Node (TQN) at EPB's Tenth Street Data Center, and the Broad Street Quantum Node (BQN) at EPB's Broad Street Data Center. Equipment for the remote nodes was configured at UTC, deployed to TQN and BQN by EPB personnel, and subsequently operated through Ethernet connections over EPB's local area network (LAN). All measurements were conducted remotely, without the need for on-site operators, reflecting a realistic field-deployment scenario. A bird’s-eye view of the deployed fiber network connectivity is shown in \textbf{Fig.~\ref{fig:networkmap}}. 

\subsection{Time-tagging module}

Photon arrival times were recorded using commercial time-tagging modules (PicoQuant MultiHarp 150P), with one module deployed at each remote node on the deployed EPB fiber network. Each module provides a timing resolution of 5~ps, a per-channel dead time of 650~ps, and an inter-channel timing jitter of 35~ps. The external 10~MHz reference input of each module was connected to the output of the co-located MAC, thereby disciplining the module's internal timebase to the corresponding rubidium frequency standard. Data acquisition was initiated remotely at UTC through the PicoQuant Python API via Ethernet connections to the EPB LAN, and the resulting timestamp streams were transferred to UTC for post-processing. Owing to the proprietary nature of the acquisition software, the exact timing of acquisition initiation is not deterministically controllable across geographically distributed nodes. Consequently, timestamp files acquired nominally simultaneously can exhibit start-time offsets of up to 1.5~s. These large and variable timing offsets constitute the primary motivation for the coincidence peak identification algorithm described in Section~4.5.

\subsection{Atomic clock initialization}

Commercial miniature rubidium atomic frequency standards (Microchip MAC-SA57) served as our connection-free timing references. Prior to each deployed network experiment, the two MACs were frequency-matched at UTC through digital frequency tuning, with agreement verified by monitoring the relative phase of their 10~MHz outputs. Following this overnight initialization, the two MACs were transported to their respective network nodes and installed for operation. Data acquisition began approximately 12~hours after the digital frequency tuning, representing a realistic deployment scenario in which clocks are prepared centrally and subsequently distributed to remote sites. After installation, the 10~MHz output of each MAC was connected to the external reference input of the co-located time-tagging module. The MACs then operated independently for the duration of the experiment, without any ongoing synchronization updates, feedback control, or communication links between nodes. The extended interval between MAC initialization and data acquisition is believed to be the dominant source of the larger accumulated timing offsets observed in the deployed network measurements compared with the local laboratory configuration.

\subsection{White Rabbit benchmarking configuration}

For all White Rabbit (WR) benchmark measurements, the time-tagging modules were referenced to a common 10~MHz clock generated by a WR switch (Safran EQP-WRS-LJ-02) located at UTC and distributed over dedicated timing fibers. Through its closed-loop synchronization architecture, the WR system continuously compensated frequency and phase deviations among the connected nodes, effectively eliminating relative timing drift and providing a benchmark synchronization performance for comparison with our connection-free MAC-based approach.
WR measurements were conducted in two configurations: (1) a local laboratory setup at UTC employing fiber delay spools (Corning SMF-28) to emulate network propagation delays, and (2) the deployed EPB fiber network with time-tagging modules located at the remote TQN and BQN sites. The local spool configuration provided a controlled baseline measurement free from the environmental perturbations present in the deployed metropolitan fiber network.


\subsection{Coincidence peak identification}

Coincidence peak identification was implemented using a two-stage search algorithm. First, temporally subsampled timestamp streams were scanned across the full range of candidate inter-file timing offsets using a relatively large 10~ns time-bin width. This coarse search efficiently identified the approximate location of the correlation peak with low yet sufficient CAR ratio, even when the acquisition start-time offset exceeded one second. Second, the full timestamp records were processed in a narrow region surrounding the coarse estimate, enabling reconstruction of the second-order cross-correlation function $G^{(2)}(\tau)$ with high resolution using a reduced time-bin width. The peak center and full width at half maximum (FWHM) were obtained from a Gaussian fit to the resulting correlation profile. The algorithm successfully recovered the coincidence peak for all laboratory and deployed network measurements.


\subsection{Phase drift correction}

To compensate for relative phase drift between the independent-running MACs, each timestamp stream was partitioned into 1~s segments, and the coincidence peak position was determined for each segment using the coarse search procedure described in Section~4.5. The resulting sequence of per-second peak positions provides a direct measurement of the accumulated timing drift throughout the acquisition. These measurements were used to construct a piecewise linear drift model, $d(t)$, with linear interpolation providing a continuous estimate of the drift over the full measurement duration. The drift model was then applied as a time-dependent correction to one of the timestamp streams. For an event recorded at time $t$, the corrected timestamp was calculated as
$t' = t - \bigl(d(t)-d(0)\bigr)$.
The subtraction of $d(0)$ preserves the absolute location of the coincidence peak, preventing an artificial shift of the reconstructed correlation function toward $t=0$. The corrected timestamp stream was subsequently processed using the fine stage reconstruction procedure described in Section~4.5 to obtain the drift-corrected $G^{(2)}(\tau)$ distribution profile and its corresponding FWHM.


\subsection{Hong-Ou-Mandel interference measurement}

Hong--Ou--Mandel (HOM) interference measurements were performed using the SPDC source in a configuration where both photons from each pair were directed to a 50:50 beam splitter (BS), with one photon traversing a motorized variable delay line (VDL) prior to interference. Coincidence counts between the two BS output ports were recorded as a function of VDL position using a 60~s integration time per position and a coincidence window of 1~ns. The HOM visibility was calculated as
\[
V = \frac{C_{\mathrm{max}}-C_{\mathrm{min}}}{C_{\mathrm{max}}},
\]
where $C_{\mathrm{max}}$ denotes the off-dip (shoulder) coincidence rate and $C_{\mathrm{min}}$ denotes the minimum coincidence rate at the center of the HOM dip \supercite{26,27}.

For all MAC-based configurations, the drift-correction procedure described in Section~4.6 was applied to the raw timestamp streams prior to HOM analysis. This ensured that the accumulated timing offsets between the two MACs did not artificially degrade the measured HOM visibility by producing a broader, less localized coincidence peak with an increased FWHM, as shown in \textbf{Fig.~\ref{fig:driftcorrection}(c)}. All experimental configurations employed the same SPDC source, optical components, and measurement procedures, with only the synchronization method and network geometry varying between measurements.




\printbibliography

\section*{Author contributions}
T.L. conceived, designed, and supervised the overall project. J.E.H., M.J.U.H., A.F.E., and I.D. performed the experiments. J.E.H. developed the computational post-processing algorithms. All authors contributed to data analysis, manuscript preparation, and revision.

\section*{Competing interests}
The authors declare no competing interests.

\section*{Acknowledgments}
J.E.H., A.F.E., I.D., and T.L. acknowledge supports from the U.S. National Science Foundation (NSF) through the ExpandQISE program under Award No. 2426699, and from the NSF CCSS program under Award No. 2503630; M.J.U.H. and T.L. also acknowledge support from the U.S. National Institute of Standards and Technology (NIST) through the CIPP program under Award No. 60NANB24D218.

\end{document}